\documentclass[fleqn,usenatbib]{mnras}

\usepackage{newtxtext,newtxmath}

\usepackage[T1]{fontenc}

\DeclareRobustCommand{\VAN}[3]{#2}
\let\VANthebibliography\thebibliography
\def\thebibliography{\DeclareRobustCommand{\VAN}[3]{##3}\VANthebibliography}

\usepackage{graphicx}	
\usepackage{amsmath}	

\title[Impact-Driven Mass Loss of Miscible Planets]{Reconciling Giant Impacts with the Radius Valley: \\
The Role of Silicate-Hydrogen Miscibility for Sub-Neptunes}

\author[E. Zlimen, H. E. Schlichting, and J. G. Rogers]{
Eva Zlimen,$^{1}$\thanks{E-mail: ezlimen@g.ucla.edu}\thanks{NSF Graduate Research Fellow}
Hilke E. Schlichting,$^{1}$
and James G. Rogers$^{2}$
\\
$^{1}$Department of Earth, Planetary, and Space Sciences, The University of California, Los Angeles, 595 Charles E. Young Drive East, Los Angeles, CA 90095, USA\\
$^{2}$Institute of Astronomy, University of Cambridge, Madingley Road, Cambridge CB3 0HA, United Kingdom\\
}

\date{Accepted XXX. Received YYY; in original form ZZZ}

\pubyear{2026}

\begin{document}
\label{firstpage}
\pagerange{\pageref{firstpage}--\pageref{lastpage}}
\maketitle

\begin{abstract}
Giant impacts are believed to play a major role in the assembly, long-term orbital stability, and resonant structure of exoplanetary systems. We revisit giant impact driven atmospheric escape while chemically coupling the interior and atmosphere by allowing for silicate-hydrogen miscibility. Incorporating hydrogen–silicate miscibility fundamentally alters atmospheric escape compared to chemically-inert models. While previous work on chemically-inert planets has found sub-Neptunes with hydrogen-dominated envelopes to undergo total atmospheric loss following collisions with impactors only a tenth of their mass, we find that including silicate-hydrogen miscibility can significantly reduce mass loss. This results from an increased envelope mean molecular weight, interior hydrogen sequestration, and modified compositional gradients in miscible planets. Miscible planets sequester hydrogen in their interiors post-impact, reducing the hydrogen available in the envelope for mass loss. As planets cool, hydrogen exsolves into their envelopes. Because more massive planets cool slower, they sequester hydrogen for longer, undergoing less mass loss. Thus, atmospheric retention depends more strongly on planet mass than impact energy, provided the impact does not gravitationally unbind the planet. Planets with masses $\gtrsim$4.50 $M_{\oplus}$ and hydrogen mass fractions \textgreater5\% typically retain atmospheres of several percent after a giant impact, while planets with masses $\lesssim$3.30 $M_{\oplus}$ are almost entirely stripped of their envelopes. This allows sub-Neptunes with hydrogen-dominated envelopes to survive giant impacts. Contrastingly, lower-mass planets, which ultimately populate the super-Earth regime, are frequently stripped of their envelopes. Our findings suggest that super-Earths and sub-Neptunes could originate from a single population with rocky interiors even if post-disk-dispersal giant impacts were common.
\end{abstract}




\section{Introduction}

The past three decades of exoplanet discoveries have been revolutionary, demonstrating a previously unimaginable diversity of planet formation outcomes. Two distinct populations of small, close-in exoplanets have emerged: super-Earths and sub-Neptunes. While the densities of super-Earths, i.e. planets smaller than about 1.6 $R_\oplus$, tend to be consistent with ``Earth-like'' compositions, sub-Neptune densities often require H/He envelopes containing a few \% of the planet's total mass \citep[e.g.,][]{Rogers2015}. Intriguingly, further observations have revealed a ``radius valley'' in the distribution of these small exoplanets. Studies have suggested that the radius valley likely marks a transition from a population of ``sub-Neptunes'' to a population of ``super-Earths''. Hydrodynamic mass loss models appear to explain the observed radius valley and related observations well. Under both photoevaporation \citep{OwenWu2013,Owen2017} and core-powered mass loss \citep{Ginzburg2018,Gupta2019,Gupta2020}, the smaller and closer-in planets lose their entire primordial atmospheres, while the larger and further away planets retain some of their envelopes. More recent work suggests that photoevaporation and core-powered mass loss may in fact have occurred in conjunction and were possibly able to enhance one another \citep{Owen2024,Misener2026}. If the radius valley was indeed carved by atmospheric escape, then the super-Earth population was born from sub-Neptunes that lost their hydrogen-dominated envelopes over time. Several observational studies find an increasing occurrence rate of super-Earths relative to sub-Neptunes over Gyr timescales, consistent with this interpretation \citep{Berger2020,David2021,Christiansen2023}.

Furthermore, super-Earths and sub-Neptunes commonly occur in multi-planet systems but often with a lack of mean motion resonances among them \citep[e.g.,][]{Fabrycky2014,Goldreich2014}. Planet formation models show that large collisions are a frequent and expected outcome needed both for long-term orbital stability after gas disk dispersal  \citep[e.g.,][]{Izidoro2017,Bean2021,Goldberg2022} and to explain the lack of mean-motion resonances in the exoplanet data \citep[e.g.,][]{Dai2024,Hansen2024,Wu2024}. However, giant impacts are difficult to reconcile with the observed ``radius valley'', as frequent collisions should have erased, or at least significantly ``polluted'', the valley by efficiently stripping sub-Neptunes of their hydrogen envelopes \citep{Biersteker2019,Biersteker2021}. This is because for planets with low mean molecular weight envelopes, giant impacts can drive mass loss both due to thermal heating from the impact and from the impact shock \citep[][]{Genda2003,Schlichting2015,Liu2015,Inamdar2016,Biersteker2019}. \citet{Biersteker2019} find that the atmospheric loss due to thermal heating dominates over the loss from the impact shock for these planets and that thermal impact-driven mass loss is highly efficient, e.g. planets will undergo total atmospheric loss following collisions with impactors only a tenth of their mass. This creates a conundrum: giant impacts appear to be an integral part of planet formation, but they are expected to populate the radius valley. Recent work by \citet{Izidoro2022} that combined breaking of resonance chains and atmospheric loss from giant impacts came to the conclusion that the radius valley can be preserved, but only if sub-Neptunes are water-ice-rich worlds and their interiors are not exclusively rocky in composition. In this scenario, super-Earths and sub-Neptunes would have to be formed as two compositionally distinct populations, which is in stark contrast to the conclusions found by photoevaporation and core-powered mass loss models which conclude that super-Earths and sub-Neptunes were born as one population with mostly rocky interiors \citep{Gupta2019,Rogers2021}.

To address these competing ideas for the origin and composition of super-Earths and sub-Neptunes, we developed the first models of atmospheric escape from giant impacts that include the physical and chemical coupling of the interior and envelope. Using the recently determined H$_2$–MgSiO$_3$ phase diagram \citep{Gilmore2026}, these new models account for the miscibility between silicate magma and hydrogen. Recent work \citep[e.g.,][]{Young2024, Rogers2025,Rogers2026} has highlighted the importance of the miscibility of silicate and hydrogen at pressures and temperatures relevant to sub-Neptunes and super-Earths. Within this new interior model, hydrogen and silicate form a completely miscible, homogeneous mixture within the interior. This allows, for example, for part of the hydrogen to mix into the silicate ``core''. Using this new interior model, we calculate the evolving planetary structure resulting from planetary-scale collisions and the associated atmospheric mass loss. We compare our results to previous work on giant impact induced mass loss that treated the rocky interior and hydrogen envelopes as two distinct and chemically-inert layers \citep[e.g.,][]{Biersteker2019, Biersteker2021}. 

The structure of the paper is as follows. In Sect. \ref{sec:methods}, we describe our new planetary structure and giant impact models and their evolution in time. In Sect. \ref{sec:results}, we present our results, highlighting the differences in mass loss and hydrogen partitioning between our new miscible and the standard, chemically-inert giant impact models. In Sect. \ref{sec:disc} we discuss implications of our results for the formation and evolution of super-Earths and sub-Neptunes along with recent observations, followed by conclusions in Sect. \ref{sec:conclusions}.

\section{Methods}
\label{sec:methods}
Throughout this work, we construct 1D, spherically symmetric sub-Neptune models and solve for hydrostatic and thermochemical equilibrium. To include miscibility self-consistently and to make the problem chemically tractable, we approximate our modelled planets as existing only of MgSiO$_3$ and H$_2$; we neglect iron. Previous works have shown that iron does not prevent hydrogen miscibility within sub-Neptunes \citep[e.g.,][]{Yuan2020, Tagawa2021, Insixiengmay2025, Young2025} and we discuss the implications of this approximation in detail in Sect. \ref{sec:iron}.

We discuss the binodal phase boundary in Sect. \ref{sec:binodal}, followed by the details of our miscible planetary structure model in Sect. \ref{sec:miscmodel} and standard, chemically-inert model in Sect. \ref{sec:immodel}. The numerical procedure used to obtain converged solutions and time evolution is presented in Sect. \ref{sec:numerics} while the giant impact model is treated in Sect. \ref{sec:impacts}.

\subsection{Phase Equilibrium: The Binodal}
\label{sec:binodal}
In this work, we use the phase equilibrium of the MgSiO$_3$-H$_2$ system as determined by \citet{Gilmore2026} to calculate the boundary, also referred to as the binodal surface, between the fully miscible, supercritical interior and the two-phase envelope. We then explore the consequences of miscibility for planetary interiors, following \citet{Rogers2025}.

To construct planetary models which account for miscibility, we must determine the phase of the material as a function of pressure, temperature, and chemical composition. Within the fully miscible interior, only one phase exists: a supercritical fluid of MgSiO$_3$-H$_2$. Conversely, the envelope is composed of a two-phase mixture of silicate-rich melt and hydrogen-rich gas. For convenience, we use the analytic fit to the binodal surface from \citet{Rogers2025} to calculate the temperature at the crest of the binodal and the mole fractions of hydrogen in the gas and melt phases, $x_{H_2}^{gas}$ and $x_{H_2}^{melt}$, respectively. As we have a two-component mixture, these imply the mole fractions of silicate as $x_{sil}^{i}$ = $1 - x_{H_2}^{i}$, where $i$ refers to the phase. 

\subsection{Chemically Interacting Planetary Structure Model}

\label{sec:miscmodel}
\begin{figure}
    \centering
    \includegraphics[trim={1.5cm 1cm 0 0},clip, width=1\linewidth]{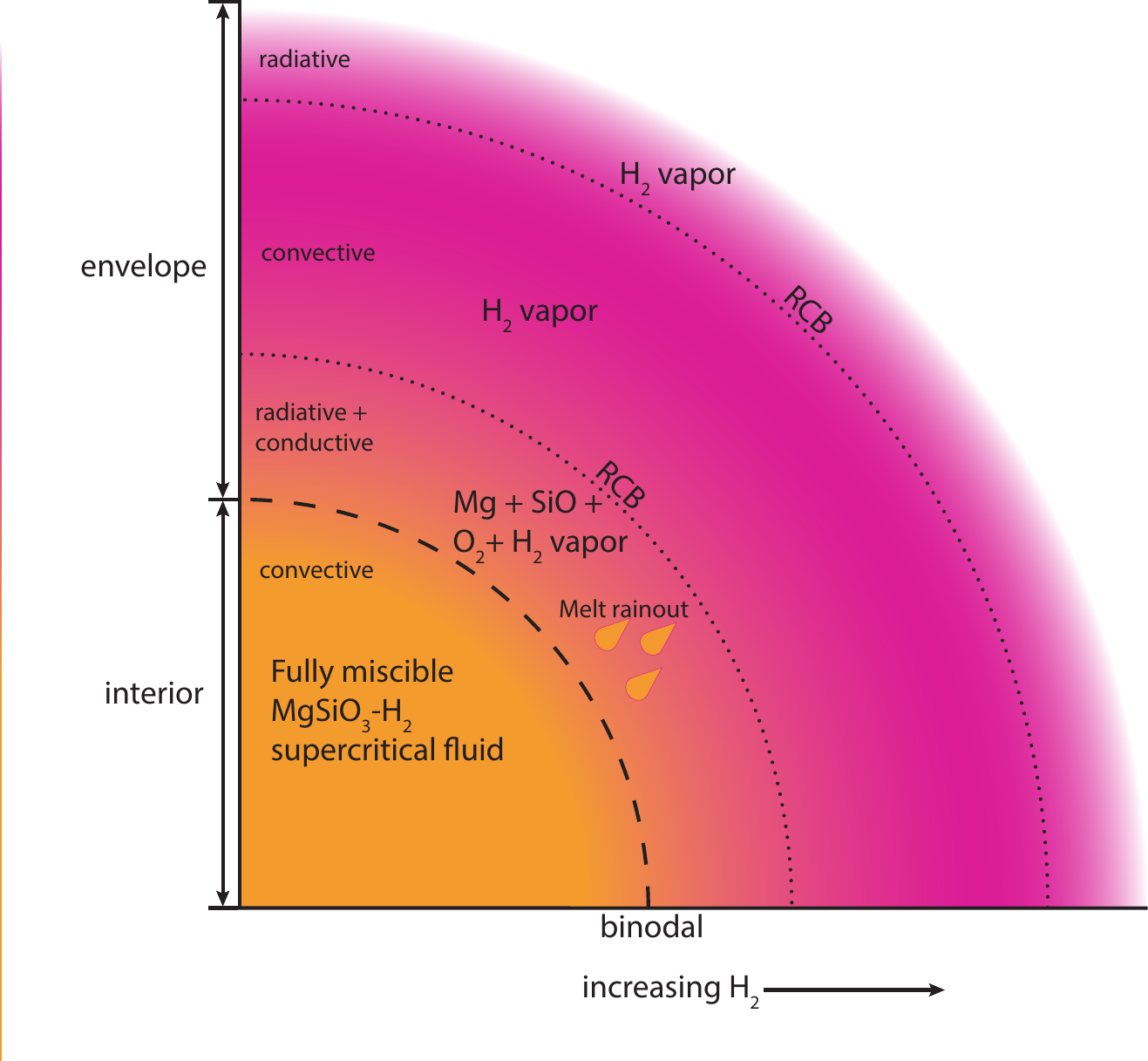}
    \caption{Schematic of our miscible sub-Neptune model. A fully miscible, supercritical MgSiO$_{3}$-H$_{2}$ interior is separated from the hydrogen-dominated envelope by the binodal phase boundary (dashed line). Directly above the binodal, convection is inhibited due the presence of silicate vapor. As temperatures and pressures decrease higher in the atmosphere, the gas becomes more hydrogen-rich and convection is no longer inhibited. A radiative upper atmosphere lies above the convective region. Radiative-convective boundaries (RCBs) are indicated by dotted lines.}
    \label{fig:modelschematic}
\end{figure}

To construct a model which allows for miscibility, we separate our planet into three regions. First, the interior, discussed in Sec. \ref{sec:interior}, is treated as a fully convective and miscible H$_2$-MgSiO$_3$ mixture bound by the binodal surface. Second, a partially convective, hydrogen-dominated envelope, discussed in Sec. \ref{sec:envelope}, sits atop the binodal boundary. Here, we allow for regions of inhibited convection due to silicate vapor in the envelope. And third, above the radiative convective boundary sits a fully radiative upper atmosphere, which completes the model. A schematic of our planetary structure model is shown in Fig. \ref{fig:modelschematic}. 

In all regions of the planet, we assume instantaneous hydrostatic equilibrium, and the planetary radius $r$, temperature $T$, and pressure $P$ for a given mass $m$ are obtained through integration of the following equations:

\begin{equation}
    \label{eq:drdmbinodal}
    \frac{dr}{dm} = \frac{1}{4 \pi r^2 \rho(r, T, P)},
\end{equation}
\begin{equation}
    \label{eq:dpdmbinodal}
    \frac{dP}{dm} = -\frac{Gm}{4 \pi r^4},
\end{equation}
\begin{equation}
    \label{eq:dTdmbinodal}
    \frac{dT}{dm} = -\frac{Gm}{4 \pi r^4} \frac{T}{P}\nabla,
\end{equation}
where $\nabla \equiv \frac{\partial lnT}{\partial lnP}$ is the temperature gradient appropriate to the region of interest.
The density $\rho(r, T, P)$ is calculated differently for each region of the planet as outlined in the following sections. Energy $E$ of the planet is calculated as 
 \begin{equation}
  \frac{dE}{dm} = -\frac{G m}{r} + \left[(1-f_{H_{2}}) C_{p, MgSiO_{3}, i}+ f_{H_{2}} C_{p, H_2}\right] T ,
 \end{equation}
 where the first term refers to the gravitational energy $E_{grav}$ and second to the thermal energy $E_{th}$. Here, $f_{H_2}$ is the hydrogen mass fraction and $C_{p}$ is the specific heat by mass where $i$ refers to the phase of $\textrm{MgSiO}_{3}$ applicable to the region of interest: melt within the interior and gas within the atmosphere. We use constant specific heats of C$_{p, H_2} = 3.5 R $, $C_{p, MgSiO_3, melt} = 120.0$ $\textrm{J} \textrm{mol}^{-1} \textrm{K}^{-1}$, and $C_{p, MgSiO_3, gas} = \frac{1}{3}(\frac{C_{p, SiO}}{\mu_{SiO}} + \frac{C_{p, Mg}}{\mu_{Mg}} +  \frac{C_{p, O_2}}{\mu_{O_2}})$ where $R$ is the gas constant and $\mu$ is the mean molecular weight. We take $C_{p, SiO} = C_{p, O_2} = 3.5 R $ and $C_{p, Mg} = 2.5R$.

\subsubsection{Interior}
\label{sec:interior}
Within the interior of the planet, which lies beneath the binodal surface, we assume a fully convective, miscible mixture of hydrogen and MgSiO$_{3}$. As a result, the hydrogen mass fraction is constant throughout the interior. Temperature and silicate density, $\rho_{sil}$, are calculated using the \citet{Wolf2018} RT Press EOS using updated constants from \citet{Luo2025}. To account for the change in density of the silicate melt due to hydrogen, we follow \citet{Rogers2025,Rogers2026} to mix the silicate density from the RT Press EOS with the hydrogen density at pressure. To calculate the temperature gradient, we assume an isentropic profile unaffected by 
the presence of hydrogen in the interior.

\subsubsection{Atmosphere}
\label{sec:envelope}
At pressures and temperatures less than that of the binodal boundary, we enter the hydrogen-dominated envelope. Here, hydrogen and silicate are no longer fully miscible and are able to exist in both liquid and vapor phases. We assume that silicate speciates into Mg, SiO, and O$_{2}$, and calculate the mean molecular weight of the gas at a given planetary radius as
\begin{equation}
\label{eq:mugas}
    \mu_{gas} = x_{H_2}^{gas} \mu_{H2} + \frac{1}{3}x_{sil}^{gas}\left(\mu_{Mg} + \mu_{SiO} + \mu_{O_2}\right).
\end{equation}
Here, $x_{sil}^{gas}$ and $x_{H_2}^{gas}$ are calculated as in Sec. \ref{sec:binodal}.
We assume an ideal gas for the envelope, and calculate the density $\rho_{gas}$ from $\mu_{gas}$ as such. We discuss the limitations of this assumption in Section \ref{sec:idealgas}. The location of the radiative-convective boundary in the upper atmosphere and the dominant mechanism of heat transport within the envelope is determined by the Schwarschild criterion. 

Within the convective region, we assume the moist adiabat of \citet{Graham2021} with complete condensate rainout; all latent heats of condensation and specific heats of components are calculated via enthalpies from the NIST database. In the upper, condensate-free region of the envelope, the moist adiabat converges to that of ideal hydrogen. Thus, we assume $\nabla_{ad, H_2, ideal} \equiv$ ($\gamma$ - 1)/$\gamma$, with $\gamma$ = 7/5. 

Additionally, we allow for convection to be inhibited within the deep envelope due to mean molecular weight gradients  \citep{markham2021, markham2022,Misener2023}. Specifically, convection is inhibited when $x_{sil}^{gas} > x_{sil, crit}^{gas}$, where the critical mole fraction of silicate in the gas phase is defined as
\begin{equation}
     x_{sil, crit}^{gas} = \left[\left(\frac{\mu_{melt} L_{cond, sil}}{R T}-1\right) \left(\frac{\mu_{gas}}{\mu_{H_2}}-1\right)\right]^{-1}.
\end{equation}
Here, $L_{cond, sil}$ is the latent heat of condensation of MgSiO$_{3}$ calculated via the NIST database, $\mu_{gas}$ is defined in Eq. \ref{eq:mugas}, and $\mu_{melt}$ is the mean molar weight of the hydrogen-silicate melt undergoing rainout:
\begin{equation}
     \mu_{melt} = x_{H_2}^{melt} \mu_{H_2} + x_{sil}^{melt} \mu_{sil}.
\end{equation}
We assume
\begin{equation}
    \label{eq:delrad}
    \nabla_{rad} = \frac{3 \kappa_{eff} P L_{RCB}}{64 \pi G m \sigma T^4},
\end{equation}
where $\sigma$ is the Stefan-Boltzmann constant, $L_{RCB}$ is the luminosity at the outermost radiative-convective boundary, and $\kappa_{eff}$ is the effective opacity of \citet{Vazan2020}:
\begin{equation}
    \kappa_{eff} = \frac{1}{{\kappa}^{-1} + {\kappa_c}^{-1}}.
\end{equation}
Here, $\kappa$ is the Rosseland mean opacity of the gas. As in \citet{Rogers2025}, $\kappa$ is calculated based on the power law optical opacity scaling for solar metallicity gas derived from \citet{Rogers2010, OwenWu2017}: 
\begin{equation}
    \kappa = 1.29 \times 10^{-2} \left(\frac{T}{1000 \mathrm{K}}\right)^{0.45} \left(\frac{P}{1 \mathrm{ bar}}\right)^{0.68} \mathrm{cm}^{2} \mathrm{g}^{-1}.
\end{equation}
Additionally, we define $\kappa_c$ as the conductive opacity, where
\begin{equation}
    \kappa_c = \frac{16 \sigma T^3}{3 \rho \lambda}
\end{equation}
and $\lambda$ is the thermal conductivity. We follow \citet{Misener2023} in using the Wiedemann-Franz law to convert experimental hydrogen electrical conductivities from \citet{McWilliams2016} to thermal conductivities. Within the upper envelope, $\kappa_{eff} \sim {\kappa}$; ${\kappa_c}$ contributes only near the binodal surface.

\subsection{Standard, Chemically-Inert Model}
\label{sec:immodel}

We compare our miscible models to standard, chemically-inert models. The standard, chemically-inert reference models are calculated in an identical way to the new miscible models except with $f_{H_2, interior}$ = 0. Similarly, the envelope is assumed to be pure hydrogen, with a fully convective deep envelope followed by an upper radiative region.

\subsection{Numerical Procedure}
\label{sec:numerics}
\subsubsection{Solving for Hydrostatic and Chemical Equilibrium}
\label{sec:static}
For each planet model, we define a planet mass $M_p$, bulk hydrogen mass fraction $f_{H_2, total}$, equilibrium temperature $T_{eq}$, outer pressure $P_{out}$, and luminosity at the outermost radiative-convective boundary $L_{RCB}$. We choose $P_{out}$ such that at minimum, $T_{RCB} > T_{eq}$. As we do not know \textit{a priori} the central conditions of the planet, we use a shooting method to obtain self-consistent solutions as in \citet{Rogers2025}. We first guess a central pressure $P_c$, central temperature $T_c$, and interior hydrogen mass fraction $f_{H_2, interior}$, and integrate from the center of the planet outwards to an intermediate boundary. 
We choose the intermediate boundary defined at $T_{inter}$ = 1500 K, at which point all silicate has rained out of the envelope. We then integrate inwards from our fixed outer boundary parameters and radius $R_{out}$ that satisfies the mass conservation implied from the first integration. 
The solution obtained from the outwards integration is then compared to the solution obtained from the inwards integration, and we iterate over our guesses of $P_c$, $f_{H_2, interior}$, and $T_c$ to obtain a self-consistent model. 

\subsubsection{Time Evolution}
\label{sec:time}
To time evolve a model from a given initial energy $E_{0}$, we solve for global energy and mass conservation using a first order Euler numerical scheme. Energy is calculated from
\begin{equation}
    \label{eq:energy}
    \frac{dE}{dt} =- L_{RCB} + \frac{G M_p \dot{m}}{R_{RCB}},
\end{equation}
where $\dot{m}$ is the mass loss rate. Here, the first term is the change in energy due to luminosity, while the second term accounts for the change in energy due to mass loss. We enforce a maximum step size in energy of 10\%. 

To account for mass loss, we assume that only hydrogen is removed from the envelope due to escape. We calculate the new planet $f_{H_2, total}$ from
\begin{equation}
\label{eq:mass}
    \frac{d f_{H_2, total}}{dt} = \frac{ \dot{m}}{M_p}.
\end{equation}

Mass loss can be driven by either photoevaporation or core-powered mass loss. Thus, we compute both rates and use the maximum at each timestep, although in reality they can enhance each other \citep{Owen2024, Misener2026}. We compute the photoevaporative mass loss rate following the analytic prescription of \citet{Misener2026}.

We determine the core-powered mass loss rate $\dot{m}_{CP}$ as the minimum of an isothermal Parker wind (as assumed in previous works, e.g., \citeauthor{Biersteker2019} \citeyear{Biersteker2019})
\begin{equation}
\label{eq:massloss}
    \dot{m_P} = - 4 \pi R_{out}^2 \rho_{out} c_{s},
\end{equation}
and the maximum mass loss rate sustained by the mass of hydrogen available in the envelope
\begin{equation}
\label{eq:mdotavail}
    \dot{m_A} = -\frac{m_{H_2, envelope}}{dt},
\end{equation}
where $R_{out}$ is the minimum of the Bondi radius R$_{B}$ and the Hill radius R$_{H}$,  $\rho_{out} = \rho (R_{out})$, and $c_{s}$ is the sound speed. We take $R_{B} = \frac{2GM_{p}}{c_s^2}$ and $R_{H} = a\left(\frac{M_{p}}{3M_{\odot}}\right)^{1/3}$. Eq. \ref{eq:mdotavail} ensures that the planet can only lose hydrogen currently available in the envelope, as opposed to that which is stored in the interior. Notably, while the timestep appears in Eq. \ref{eq:mdotavail}, $\dot{m_A}$ is only weakly timestep dependent. This is because $m_{H_2, envelope}$ also depends on the timestep: for a longer timestep, the planet cools further. Thus, more hydrogen can exsolve from the envelope, increasing $m_{H_2, envelope}$. 

We emphasize that in order to continually lose mass, the mass at the outer radius must be continually replenished. Thus, we must account for the energy required to resupply mass to $R_{out}$. We note that we operate in a regime different from the one than discussed by \citet{Modirrousta2026}, who assume that the envelope is optically thick up to the Bondi radius. Thus, to properly account for this resupply of material, we take into account two energy limits for core-powered mass loss. The first is the energy required to lift material from the RCB to the photospheric radius $R_{ph}$. This energy is provided by the cooling luminosity of the planet at the radiative-convective boundary, $L_{RCB}$, and yields $\dot{m}_{cool}$. The second is the energy required to lift material from $R_{ph}$ to $R_{out}$. This is provided by the stellar bolometric luminosity and yields $\dot{m}_{bol}$. The planet's cooling luminosity no longer contributes in this region as it is optically thin.

Accordingly, to calculate the energy-limited core-powered mass loss rate $\dot{m}_{EL}$, we take the minimum of  $\dot{m}_{cool}$ and $\dot{m}_{bol}$. We use the minimum of $\dot{m}_{EL}$ and $\dot{m}_{CP}$ as the true core-powered mass loss rate. We find that mass loss from our miscible models is largely controlled by $\dot{m_A}$ immediately after a giant impact. We use a minimum $dt$ set by the hydrogen exsolution timescale ($10^{6}$ seconds), motivated by the convective timescale of a 4 $M_{\oplus}$ body as derived through mixing length theory \citep[e.g.,][]{Guillot2004}. 
 
\subsection{Giant Impacts}
\label{sec:impacts}
We simulate giant impacts between a target body of mass $M_{t}$ and impactor with mass $m_{imp}$ as follows. We assume an initial cooling timescale $t_c = E_{th}/L_{RCB}$ = 10$^{8}$ years to set the initial radius and energy budget of the target. We then evolve the target planet for 10$^{9}$ years before the impact occurs. As the cooling timescale is short compared to the planet's age at the time of impact, the influence of the exact initial cooling timescale is negligible. In addition, we do test the sensitivity of our mass loss results to different planet ages at the time of impact in Sect. \ref{sec:agedep}. 

At the time of impact, we add the impactor and target mass and assume all the impact energy goes into heating the interior of the newly formed planet. We approximate the impact energy as $E_{imp} = \frac{1}{2} m_{imp} v_{imp}^2$. We take $v_{imp}$ as the mutual escape velocity of the planet, enhanced by a factor $C$ (set to 1 unless otherwise specified):
\begin{equation}
    v_{imp} = C\left(\frac{2 G M_{t}^2}{R_{t} (M_{t} + m_{imp})}\right)^{1/2},
\end{equation}
where we approximate the impact radius as $R_{t}$, the radius of the target body. We evolve the energy and mass of the planet immediately after the impact according to Eqs. \ref{eq:energy} and \ref{eq:mass} for 10$^{10}$ years.

For convenience, we assume both the target and impactor bodies have the same hydrogen mass fraction, equal to that of the post-impact planet. However, our results depend only on the planet mass and hydrogen content immediately after the impact and therefore encompass many different combinations of possible impactor and target scenarios.

\section{Results}
\label{sec:results}

In the following subsections, we present the giant impact driven atmospheric mass loss results for fully miscible planets. We begin with an illustrative example in Sect. \ref{sec:example}, comparing mass loss of a fully miscible model to that of a standard, chemically-inert model. We then consider the dependence of atmospheric mass loss on planet mass and age in Sects. \ref{sec:massdep} and \ref{sec:agedep}, respectively. We finish by examining the effect of initial atmospheric mass fraction on the extent of atmospheric loss in Sect. \ref{sec:massfrac}. 

\subsection{Comparing the Standard and Miscible Mass Loss Case: An Illustrative Example}

\label{sec:example}
\begin{figure*}
 \centering
    \includegraphics[trim={0 0 0 1.5cm},clip,width=\linewidth]{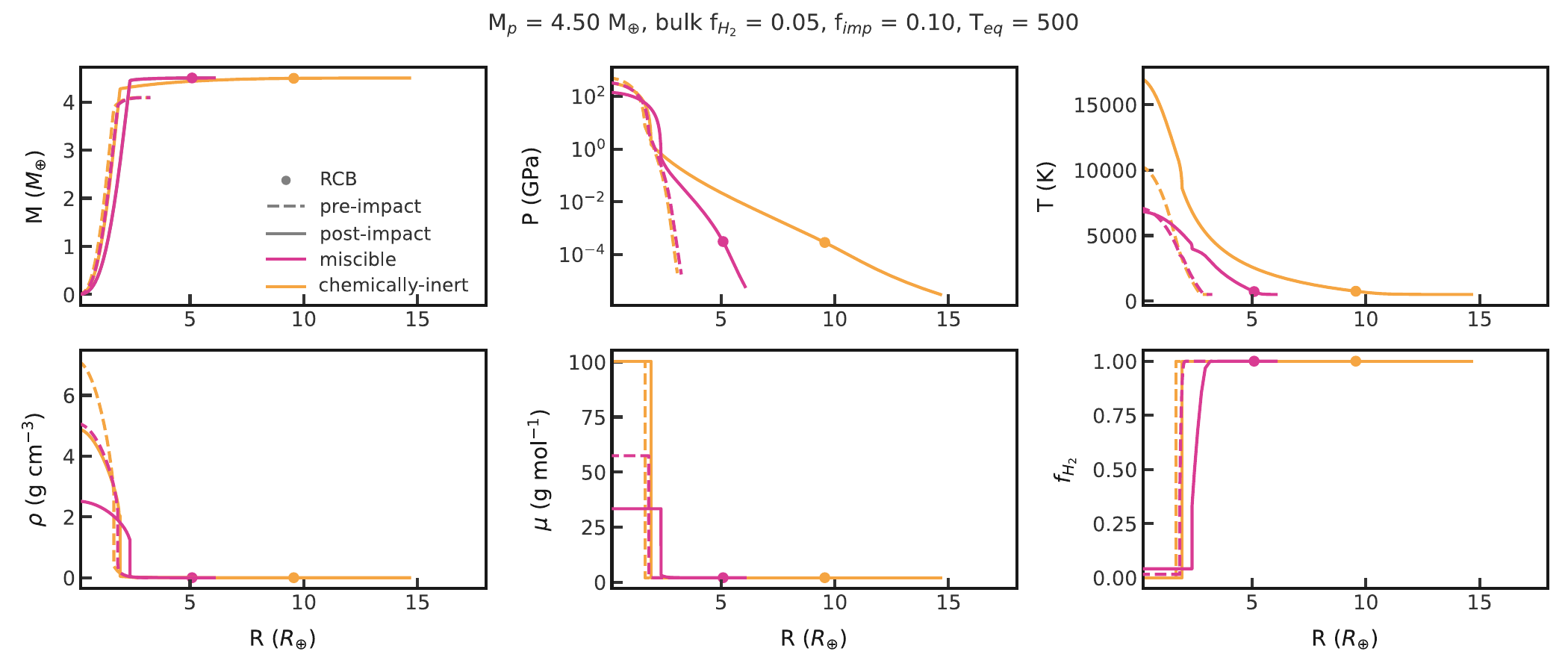}
    \caption{Interior structure of a 4.09 $M_{\oplus}$ miscible (pink) and chemically-inert (orange) target planet  before (dashed) and after (solid) an impact to yield a 4.50 $M_{\oplus}$ planet. Mass, pressure, temperature, density $\rho$, mean molar weight $\mu$, and hydrogen mass fraction $f_{H_2}$ are shown vs radius, in $R_{\oplus}$, out to the photospheric radius. The post-impact RCB is indicated by filled circles. Following an impact, the chemically-inert planet is inflated to a larger radius without greatly altering the size of the interior, while the miscible planet has a more extended interior. The density (lower left panel) of the miscible planet interior is greatly reduced post-impact due to an increase of hydrogen within it. As seen in the bottom right panel, the miscible planet has a more gradual compositional gradient in hydrogen content than the chemically-inert one as it transitions from a silicate-dominated interior to hydrogen-dominated envelope.}
    \label{fig:structure}
    
\end{figure*}

\begin{figure*}
    \centering
    \includegraphics[trim={0 0 0 1.2cm},clip, width=\linewidth]{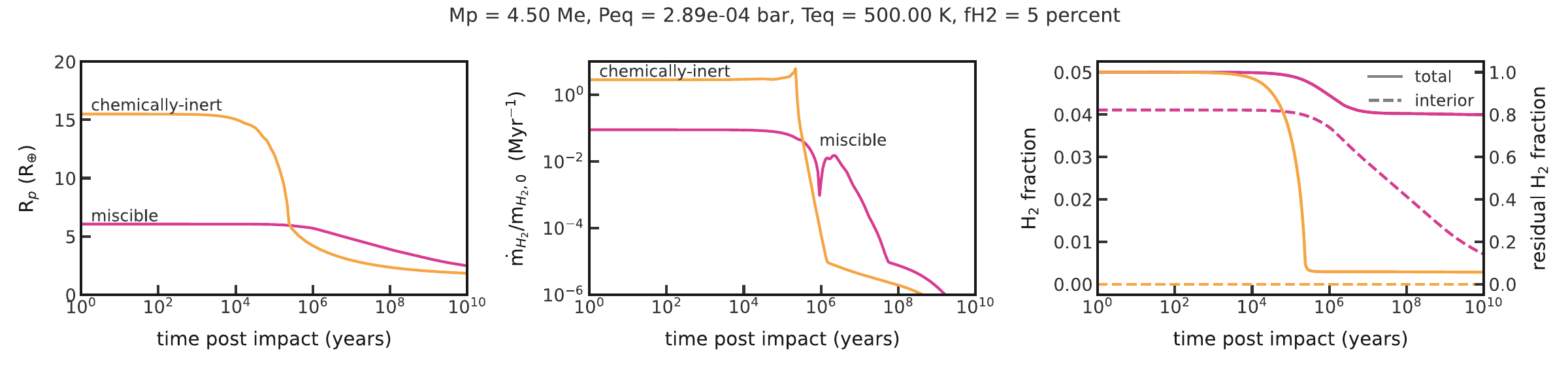}
    \caption{Post-impact time evolution of a 4.50 $M_{\oplus}$ planet with an equilibrium temperature of 500 K formed from a collision between a 4.09 $M_{\oplus}$ target planet and 0.41 $M_{\oplus}$ impactor for both a miscible (pink) and chemically-inert (orange) case. Shown in the first column is the time evolution of the photospheric radius R$_{p}$ while in the second column is the mass loss rate as a fraction of the initial hydrogen mass. In the third column is the time evolution of the interior hydrogen mass fraction (dashed) and total planet hydrogen mass fraction (solid). The fraction of hydrogen remaining is indicated on the right axis.  Immediately post-impact, the miscible planet sequesters the majority of its hydrogen in the interior, thus preventing extreme core-powered mass loss. In contrast, the standard, chemically-inert model has twice the radius of the miscible model and several orders of magnitude higher mass loss rates post-impact. Note that the mass loss rate of the miscible model sharply decreases at $\sim10^{6}$ years. Here, mass loss is limited by the rate at which hydrogen can exsolve from the interior. This is due to depletion of hydrogen within the envelope. Once sufficient hydrogen has exsolved from the interior, mass loss can proceed more rapidly. 
}
    \label{fig:time}
\end{figure*}

 \begin{figure*}
    \centering
    \includegraphics[trim={0 .05cm 0 0},clip, width=\linewidth]{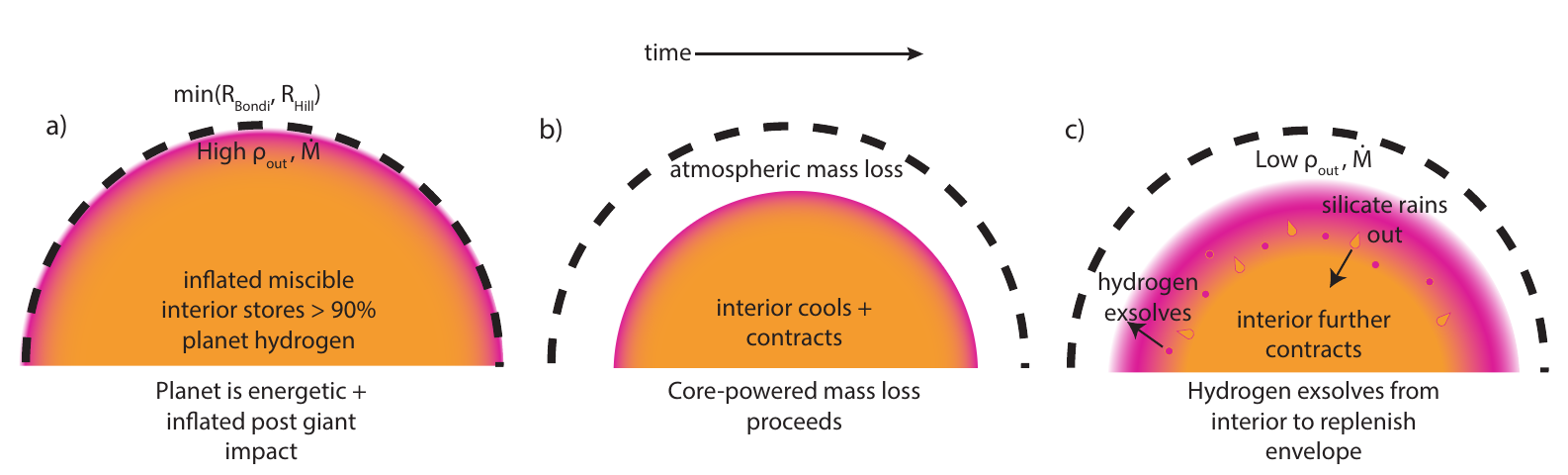}
    \caption{Schematic of hydrogen exsolving from the interior of a miscible planet following a giant impact. In column a), immediately post-impact, the planet has an inflated silicate-dominated interior (orange) storing the majority of its hydrogen. A small hydrogen-dominated envelope (pink) sits atop the interior. While this envelope has a high density at the outer radius of the planet (black dashes), defined by the minimum of Bondi and Hill radii, the mass of the envelope itself is small. As core-powered mass loss proceeds through column b), the small available envelope is lost as the interior cools. Both processes drive the radius of the planet to shrink. This new, less energetic state is shown in column c), where hydrogen can now exsolve from the interior to replenish the envelope. At this point, the planet is less inflated than it was immediately post-impact, allowing for lower densities at the outer radius and consequently a lower mass loss rate. }
    \label{fig:hexsolve}
\end{figure*}

To understand the behavior of miscible planets, we first directly compare the post-impact mass loss of a fully miscible model to that of a standard, chemically-inert model for a fiducial 4.50 $M_{\oplus}$ planet. We assume this planet initially has a 5\% hydrogen envelope and was formed from a collision between a 4.09 $M_{\oplus}$ target planet and 0.41 $M_{\oplus}$ impactor. We assume an equilibrium temperature of 500 K, corresponding to a semimajor axis of about 0.26 AU around a Sun-like star.

\subsubsection{Comparing the Post-Impact Interior Structure of Miscible and Chemically-Inert Planets}

The difference in interior structures immediately before and after a giant impact for both a miscible planet and standard, chemically-inert planet are shown in Fig. \ref{fig:structure}. Prior to the impact, the miscible model has a reduced interior density when compared to the standard, chemically-inert model. This is due to hydrogen 
mixed into the interior. However, both models have comparable pre-impact radii. Post-impact, the two models become increasingly dissimilar: the standard, chemically-inert planet has a radius twice that of the miscible planet.  

The differing post-impact radii of the miscible and chemically-inert models can be understood as a combination of three factors: (i) the conditions at the interior-atmosphere boundary, (ii) hydrogen incorporation into the interior, and (iii) the mean molecular weight of the atmosphere. For miscible models, the temperature at the interior-atmosphere interface is given by that of the binodal phase boundary. For the MgSiO$_{3}$-H$_{2}$ system and pressures relevant to sub-Neptune planet masses, this temperature is $\sim$ 4000 K \citep[][]{Gilmore2026}. In the chemically-inert case, there is no such constraint. As a result, the temperature at the interior-envelope boundary post-impact is twice that of the miscible model, as dictated by the impact energy, resulting in larger envelope radii. 

The second process, hydrogen dissolution into the interior, is only allowed within the miscible model. We see directly in the lower right panel of Fig. \ref{fig:structure} that the mass fraction of hydrogen in the miscible model is elevated within the interior post-impact. This decreases the interior density and mean molecular weight of the interior in the miscible model by a factor of two compared to pre-impact values. As the central temperature remains relatively unchanged, we can attribute this drop in density to the higher fraction of hydrogen sequestered within the interior post-impact. Thus, for miscible models, increased hydrogen abundances within the interior contributes to a smaller post-impact radius when compared to chemically-inert models.

Additionally, the miscible model allows for silicate vapor in the atmosphere. Post-impact, we see an increase in atmospheric density and mean molecular weight at the base of the atmosphere due to these higher mean molecular weight species. This limits the post-impact increase in radius for miscible planets. Further, the addition of silicate to the atmosphere allows for regions in the envelope where convection is inhibited, contributing to increased cooling timescales for miscible models. 

In contrast, the chemically-inert model does not allow for hydrogen to exist within the interior nor for silicate species in the atmosphere. For the chemically-inert model, the post-impact increase in the envelope radius is driven by the impact energy increasing the temperature at the interior-atmosphere boundary. Thus, for the same impactor, the radius of the chemically-inert planet increases more than that of the miscible one.

\subsubsection{Atmospheric Mass Loss}
When planets undergo thermally driven atmospheric escape, mass is stripped from the planet's outer boundary, defined as the minimum of the Bondi and Hill radii. Since the upper radiative regions of the envelopes are approximately isothermal, their densities fall off exponentially above the radiative-convective boundary. As a result, the mass loss rate is highly sensitive to the size of a planet's radiative-convective boundary. Therefore, as standard, chemically-inert models have larger post-impact radiative-convective boundaries than miscible models, we expect fully miscible planets to undergo less mass loss than chemically-inert ones.

And indeed, we find miscible planets have lower mass loss rates after giant impacts, as shown in Fig. \ref{fig:time}. Post-impact, the standard, chemically-inert planet has twice the radius of the miscible model, and $\sim 100$ times the mass loss rate. As a result, the chemically-inert model loses its envelope before 10$^6$ years of evolution. 

Conversely, the miscible model retains an atmosphere over 10 billion years. The miscible model sequesters much of its hydrogen within the interior and so undergoes little mass loss within the first 10$^6$ years. However, the remaining small hydrogen atmosphere is depleted during this time, shown by a drop in the mass loss rate around 10$^6$ years. More hydrogen must exsolve from the interior to allow for continued mass loss. As the miscible model cools and contracts, hydrogen can exsolve from the interior to replenish the envelope. However, the mass loss rate decreases as less energy is available to drive mass loss.

Thus, for miscible models, not only does the mass loss rate evolve with time, but so does the amount of hydrogen stored within the interior. The interplay between hydrogen exsolution and mass loss for a miscible planet is illustrated in detail in the right column of Fig.  \ref{fig:time} and schematically in Fig. \ref{fig:hexsolve}. Post-impact, the interior grows as the majority of the hydrogen mixes into the interior. This leaves only a thin hydrogen-dominated atmosphere. Due to the addition of the impact energy, the planet has a greater radius when compared to its pre-impact state. Initially, the thin hydrogen envelope is lost in the miscible model, but this is only a small fraction of the planet's total hydrogen budget and the hydrogen stored within the interior is retained. As discussed in Sect. \ref{sec:binodal}, the binodal phase boundary depends on the planetary temperature and pressure profiles, along with the hydrogen mass fraction. After cooling both radiatively and through mass loss, the planet has a slightly lower temperature at the binodal surface. This produces a phase equilibria which allows for a higher fraction of hydrogen to exist as a gas, as opposed to within the supercritical miscible interior. As a result, hydrogen exsolves from the interior into the atmosphere. During this process, the planet continues to cool radiatively and to contract, and so the density at the outer radius is reduced and hence its mass loss rate is decreased.
 
As a result, the miscible planet is able to retain $\sim$80\% of its original hydrogen. In the standard, chemically-inert case, no hydrogen can be sequestered into the interior, leading to a more inflated radius and higher mass loss rate post-impact. Thus, for standard chemically-inert models, a 10\% mass fraction impactor causes complete mass loss in under a million years. Now that we have developed an intuition and physical understanding of the mass loss evolution of miscible planets, we turn to investigating a larger parameter space.

\subsection{Dependence of Atmospheric Loss on Planet Mass and Impact Energy}

\label{sec:massdep}
\begin{figure}
    \centering
    \includegraphics[trim={11.65cm .5cm 0cm 2.21cm},clip,width=\linewidth]{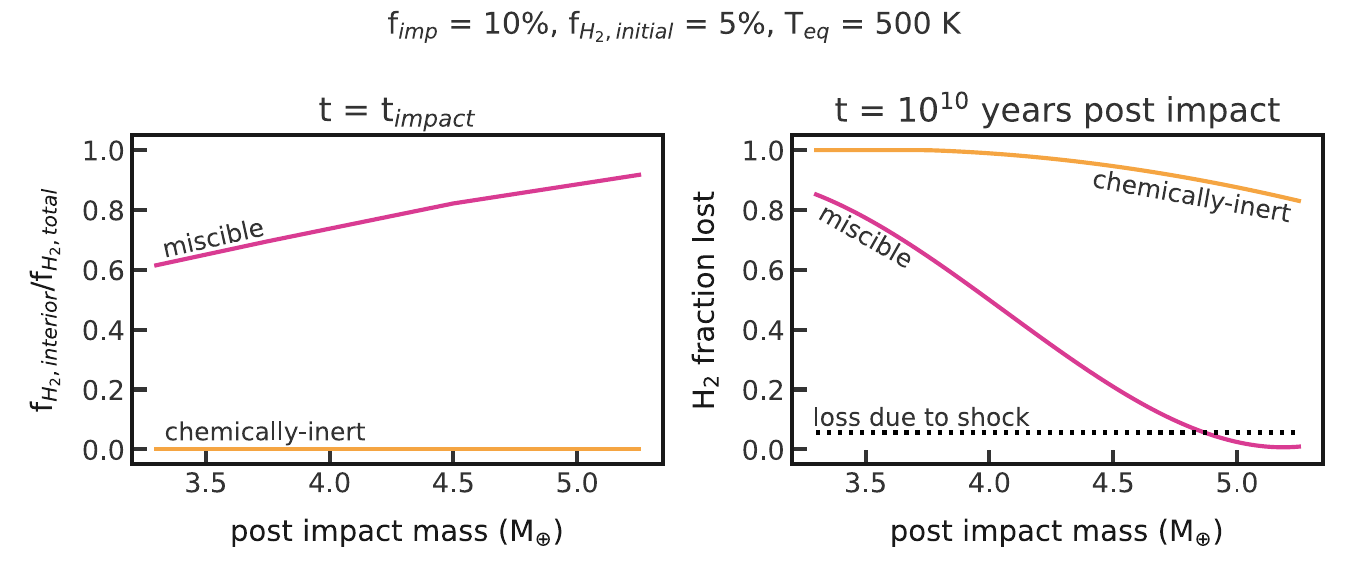}
    \caption{Total hydrogen fraction lost over time vs post-impact planet mass for chemically-inert (orange) and miscible (pink) models following impacts with impactors of 0.1$M_{t}$, where $M_{t}$ is the mass of the target planet. Loss due to the impact shock is shown as the black dotted line (Eq. 33 of \citealt{Schlichting2015}), which is only important for higher mass miscible planets. Planets are assumed to initially have 5\% hydrogen envelopes. As planet mass increases, the amount of hydrogen lost is reduced substantially in the miscible case. Higher mass miscible planets undergo negligible mass loss.}
    \label{fig:multimasscomp}
\end{figure}

\begin{figure}
    \centering
    \includegraphics[trim={0 .5cm 11.5cm 2.21cm},clip,width=\linewidth]{multimasscompare222polished.pdf}
    \caption{Fraction of hydrogen stored in the interior immediately post-impact  vs post-impact planet mass for chemically-inert (orange) and miscible (pink) models following impacts with impactors of 0.1$M_{t}$, where $M_{t}$ is the mass of the target planet. Planets are assumed to initially have 5\% hydrogen envelopes. For miscible planets, planet mass controls how much hydrogen can be stored in the interior immediately post-impact. More massive miscible planets can sequester more hydrogen within the interior, and so undergo less mass loss.}
    \label{fig:multimasscompint}
\end{figure}

We now investigate how planet mass affects atmospheric mass loss in both the miscible and the standard, chemically-inert case. Within this section, we model all planets as initially having 5\% hydrogen envelopes and equilibrium temperatures of 500 K, corresponding to a semimajor axis of $\sim$0.26 AU around a Sun-like star. We span post-impact masses between 3.30 $M_{\oplus}$ and 5.25 $M_{\oplus}$, formed from collisions by impactors 10\% the mass of the target body. We also investigate a range of impact energies by varying the impact velocity within a factor of two of the mutual escape velocity

Intuitively, we would expect higher mass planets to undergo less mass loss due to their greater gravitational binding energy. However, for miscible models, we find the hydrogen fraction lost to be inversely proportional to the planet mass due to hydrogen sequestration increasing for more massive planets. As shown in Fig. \ref{fig:multimasscomp}, the dependence of mass loss on planet mass is seen more strongly for miscible models than for standard, chemically-inert models. Additionally, for all considered chemically-inert planet masses, impact-induced mass loss removes the majority of the envelope. Contrastingly, miscible planets over $\sim$4$M_{\oplus}$ retain at least half of their envelopes after the giant impacts considered here.

These finding are a direct result of how, for miscible planets, planet mass affects the amount of hydrogen stored within the interior. The composition of the supercritical interior, as discussed in Sect. \ref{sec:binodal}, depends on the amount of hydrogen able to exist as a gas in the atmosphere. By mass conservation, all hydrogen that is not within the envelope must be stored within the interior. The mass fraction of hydrogen in the gas phase is calculated from the temperature and pressure at each radial shell in the atmosphere. At higher temperatures, the gas becomes increasingly silicate-rich. This hydrogen ``deficit'' within the gas is balanced by increased hydrogen within the interior. 

As higher mass planets have higher central temperatures and pressures, they also have higher temperatures at the binodal and in the silicate rainout region just above. Near the crest of the binodal surface, for a given pressure, the compositions of the melt and gas become highly sensitive to changes in temperature \citep[e.g.,][]{Young2024}. As a result, small differences in binodal temperatures, driven by differences in planet mass, greatly impact the amount of hydrogen within the gas phase. Thus, as shown in Fig. \ref{fig:multimasscompint}, more massive planets sequester more hydrogen in the interior. Correspondingly, they have less hydrogen within their atmospheres immediately post-impact.

\begin{figure*}
    \centering
    \includegraphics[width=\linewidth]{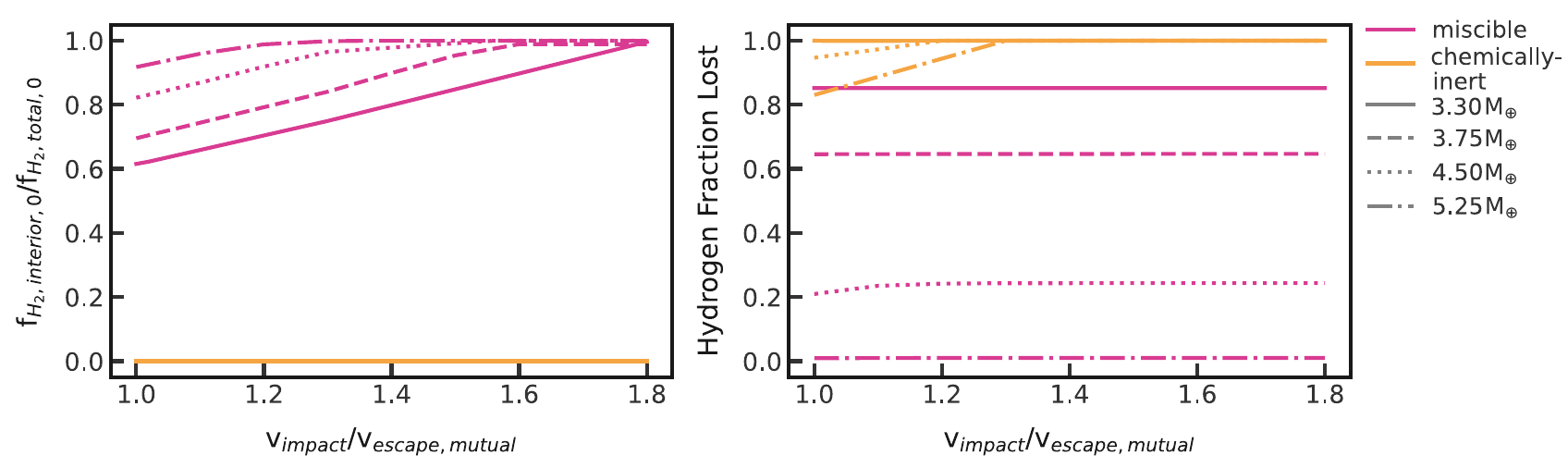}
    \caption{Fraction of hydrogen stored in the interior immediately post-impact (left) and total hydrogen fraction lost over time (right) versus impact velocity as a fractional increase of the mutual escape velocity for miscible (pink) and chemically-inert (orange) planets of different masses. Planets are assumed to initially have 5\% hydrogen envelopes. 
    For miscible planets, mass loss is reduced as planet mass increases. While increasing the impact velocity increases the amount of hydrogen stored in the interior immediately post-impact, little difference is seen in mass loss between different impact velocities. Further increases in impact velocity do not affect the hydrogen fraction lost.}
    \label{fig:eimpactcomp}
\end{figure*}

Additionally, the exact impact energy has a weaker effect than the planet mass itself on the extent of mass loss. This is shown in Fig. \ref{fig:eimpactcomp}. Here, we vary the impact energy by increasing the impact velocity. Physically, we can interpret this enhanced impact velocity as the impactor coming in with an additional relative velocity such that it is accelerated above the mutual escape velocity upon impact. The amount of mass loss from miscible models is largely unchanged by increases in impact velocity and impact energy. Standard, chemically-inert models are also relatively unaffected; however, this is due to the fact that the chemically-inert planet models were already experiencing near-total atmospheric loss for impact velocities equal to the mutual escape velocity.  For miscible planets, the impact energy controls the amount of hydrogen stored in the interior immediately post-impact, as seen in the left panel of Fig. \ref{fig:eimpactcomp}. More energetic impacts increase the temperature near the binodal surface. 
As discussed above, this allows the planet to store more hydrogen within the interior. 

\begin{figure*}
    \centering
    \includegraphics[trim={0 0cm 0 0cm},clip,width=\linewidth]{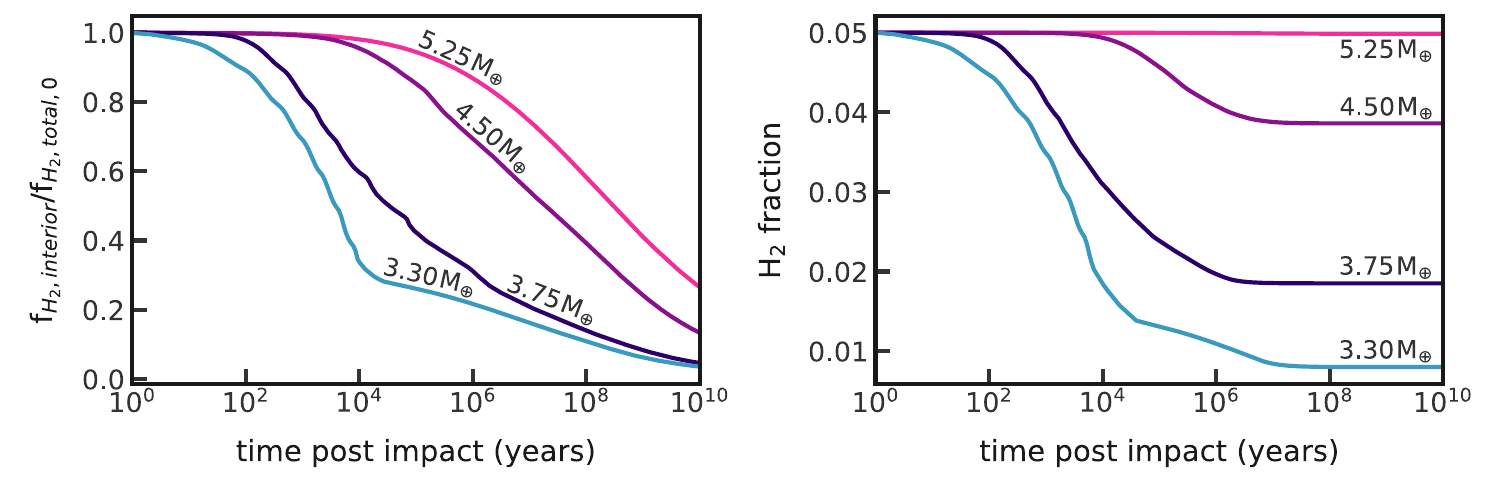}
    \caption{Fraction of hydrogen stored in the interior normalized to the initial total planet hydrogen mass fraction (left) and the total planet hydrogen mass fraction (right) vs time post-impact for four miscible planet masses. Planets are assumed to initially have 5\% hydrogen envelopes. Post-impact energies were chosen such that \textgreater 95\% of the hydrogen was sequestered in the interior immediately post-impact. Higher mass planets store more hydrogen in the interior over longer times, preventing mass loss from occurring as rapidly. Lower mass planets exsolve their hydrogen more quickly and lose more of their envelopes.}
    \label{fig:htime}
\end{figure*}

However, as shown in the right panel of Fig. \ref{fig:eimpactcomp}, simply storing more hydrogen in the interior immediately post-impact does not translate to less mass lost over time. This is because it is the planet mass that controls the speed at which the hydrogen exsolves from the interior, as shown in Fig. \ref{fig:htime}. The dependence on planet mass arises from the fact that higher mass planets cool more slowly. The relationship between the cooling time and planet mass is $\propto r \propto m^{1/3}$, i.e. the ratio of the mass of material that needs to cool ($\propto r^{3}$) to the surface area over which heat is radiated ($\propto r^{2}$). Therefore, as lower-mass planets cool after an impact, their phase equilibria will allow for more hydrogen to exist in the gas phase at earlier times. The envelope is then more hydrogen-rich shortly following an impact, when the planet is more energetic and has a larger radius. Thus, these less massive planets undergo greater mass loss, independent of how much hydrogen is sequestered within the interior immediately post-impact. More massive miscible planets undergo less mass loss because hydrogen exsolution is not as rapid.

\subsection{Dependence of Atmospheric Mass Loss on Planet Age}

\label{sec:agedep} 
\begin{figure}
    \centering
    \includegraphics[trim={0cm 0cm 0cm 0cm},clip,width=\linewidth]{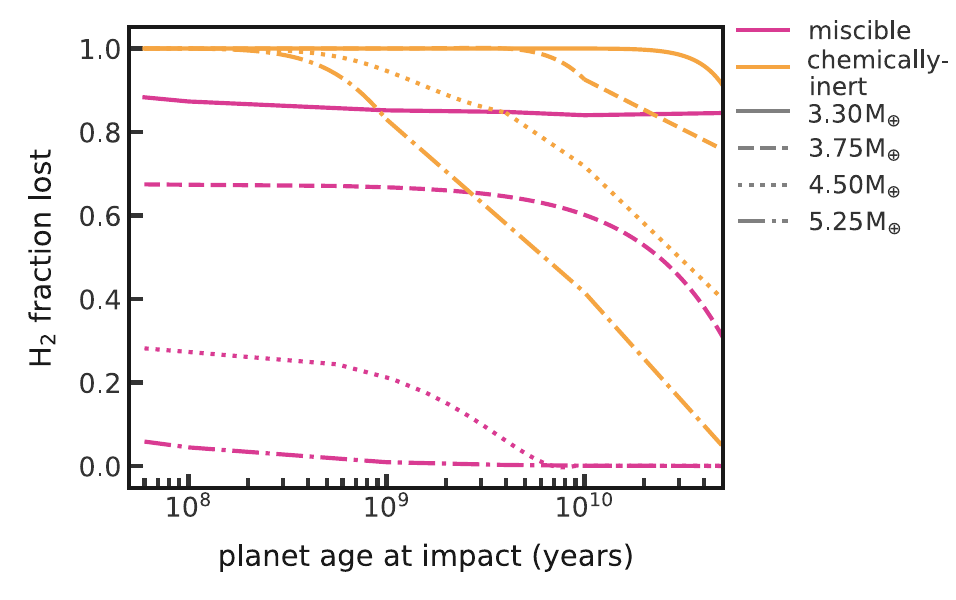}
    \caption{Total hydrogen fraction lost over time as a function of planet age at the time of impact (years) for various masses of miscible (pink) and chemically-inert (orange) planets. Planets are assumed to initially have 5\% hydrogen envelopes.  Regardless of the exact planet age, miscible planets retain a substantially larger envelope fraction than chemically-inert models for planet ages less than $\sim10^{10}$ years.} 
    \label{fig:agedep}
\end{figure}

In the previous sections, we simulated a Gyr-old planet undergoing a late-stage giant impact. Here, we relax this assumption and investigate the effect of planet age on the extent of mass loss. Following the logic of Sect. \ref{sec:example}, we would intuit that a less energetic, cooler planet would undergo less mass loss after an impact. This corresponds to an older planet age at the time of impact. We find this intuition to be true for both miscible and standard, chemically-inert models, as shown in Fig. \ref{fig:agedep}. However, miscible models are less sensitive to the planet age than chemically-inert models. For miscible models, planet mass remains the dominant factor in determining mass loss. This follows from the strong dependence of planet mass on the exsolution timescale.

For a given planet mass, the mass loss underwent by miscible and standard, chemically-inert models becomes more similar as the planet age at impact increases. We can understand this from the dependence of miscibility on temperature. In order to sustain a fully miscible interior, planets must have high enough temperatures and pressures to trigger a phase change. As miscible planets cool, their fully miscible region becomes smaller. As a result, cooler miscible planets more closely resemble chemically-inert planets, and thus the two models have more similar evolutions. However, the two models produce distinct mass loss outcomes for impact times up to $\sim$ 10 billion years. While planet age partially controls the extent of mass loss, it does not erase differences in evolution due to the presence of  miscibility. 

Additionally, we note that our results are unaffected if we assume that the pre-impact target body was cold and chemically-inert. We verify that the impact energy of our chemically-inert models, when applied to a miscible model, is sufficient to produce a fully miscible interior. Hence, the mass loss results presented here are not sensitive to the exact formation history of the planet before impact.

\subsection{Dependence of Atmospheric Mass Loss on Initial Hydrogen Mass Fraction}

\label{sec:massfrac}
\begin{figure*}
    \centering
    \includegraphics[trim={0cm 0cm 0cm 0cm},clip,width=\linewidth]{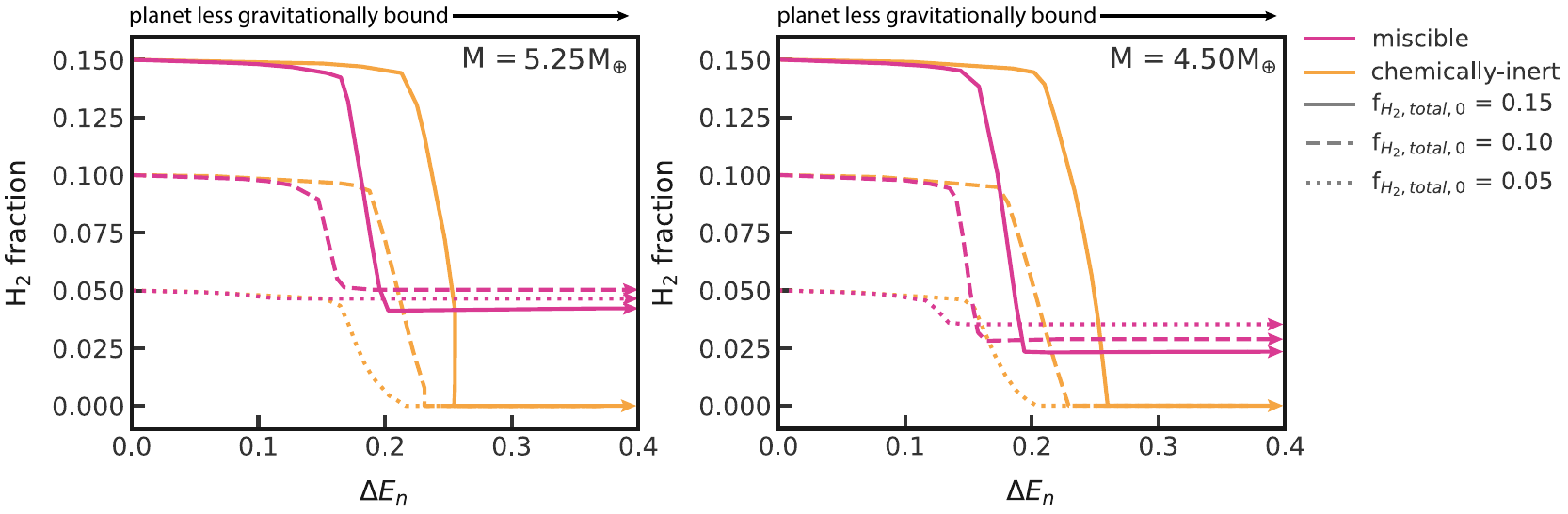}
    \caption{Total hydrogen mass fraction remaining over time for a 5.25 $M_{\oplus}$ planet (left) and a 4.50 $M_{\oplus}$ planet (right) versus the normalized energy $\Delta E_{n}$, defined as in Eq. \ref{eq:enorm},
    for three different initial hydrogen mass fractions (f$_{H_2, total, 0}$). Planet masses are given as the post-impact mass. Standard, chemically-inert models (orange) lose their entire envelopes for $\Delta E_{n} > 0.3$. Miscible models (pink) retain, at a minimum, a few percent hydrogen envelope for all $\Delta E_{n} < 1$. Mass loss does not increase above $\Delta E_{n}\sim0.4$, until the planet is fully gravitationally unbound post-impact ($\Delta E_{n} > 1$).}
    \label{fig:hlostemf}
\end{figure*}

Finally, we expand our analysis to include planets with initial hydrogen mass fractions of 10\% and 15\% to more completely span the sub-Neptune regime. We compare to our previous findings for planets with initial hydrogen mass fractions of 5\%.
For convenience, we define a dimensionless ``normalized energy'' as
\begin{equation}
\label{eq:enorm}
    \Delta E_{n} = 1- \frac{E_{total}}{E_{grav}}\bigg|_{t = t_{impact}},
\end{equation}
where $E_{total}$ is the sum of the planet's gravitational and thermal energy, and this value is calculated immediately after the impact. For a bound planet, both $E_{total}$ and $E_{grav}$ are negative and $|E_{grav}| \geq |E_{total}|$; as a result, $\Delta E_{n}$ is always between 0 and 1. A normalized energy of 0 would correspond to a planet with only gravitational energy. As the thermal energy increases and approaches the magnitude of the gravitational energy, the total energy will go to zero. This corresponds to a marginally gravitationally bound planet with $\Delta E_{n} = 1$. Once the magnitude of the thermal energy is greater than that of the gravitational energy, the total energy becomes positive, and the planet is unbound ($\Delta E_{n} > 1$). In this sense, we can think of $\Delta E_{n}$ as a measure of how far away the planet is from being gravitationally bound.

We show the total hydrogen fraction remaining post-impact as a function of $\Delta E_{n}$ for two planet masses in Fig. \ref{fig:hlostemf}. Planet masses are given as the sum of the target planet and impactor masses. We find that the mass loss of both chemically-inert and miscible models are relatively insensitive to changes in initial hydrogen mass fraction. Higher hydrogen mass fractions require a larger $\Delta E_{n}$ to undergo mass loss, consistent with previous works \citep[e.g.,][]{Biersteker2019}. As $\Delta E_{n}$ increases, the hydrogen fraction remaining after impact converges to a narrow range of values across all initial mass fractions. 

We see that as $\Delta E_{n}$ first begins to increase, miscible models undergo more mass loss than standard, chemically-inert models. This is because mass loss begins at a lower $\Delta E_{n}$ for miscible planets than for chemically-inert planets. We can understand this behavior as follows. As previously discussed, miscible planets are less centrally condensed due to hydrogen sequestration within the interior. As a result, miscible models have a lower $|E_{grav}|$  and, correspondingly, $\Delta E_{n}$, than chemically-inert models of the same total energy. This drives mass loss at lower normalized energies for miscible models than standard, chemically-inert models.

Critically, as $\Delta E_{n}$ increases above $\sim$0.3, all chemically-inert models are entirely stripped of their envelopes. Contrastingly, for the same $\Delta E_{n}$, miscible planets retain a $\sim3-5\%$ envelope regardless of initial hydrogen mass fraction. Again, we find that planet mass is the most important factor in determining the extent of mass loss of miscible models. Giant impacts which strip the envelopes of standard, chemically-inert planets do not cause total atmospheric mass loss of miscible planets, independent of the initial hydrogen mass fraction. 

\section{Discussion}
\label{sec:disc}

\subsection{Implications for the Observed Exoplanet Population}

\begin{figure}
    \centering
    \includegraphics[trim={0cm 0cm 0cm 0cm},clip,width=\linewidth]{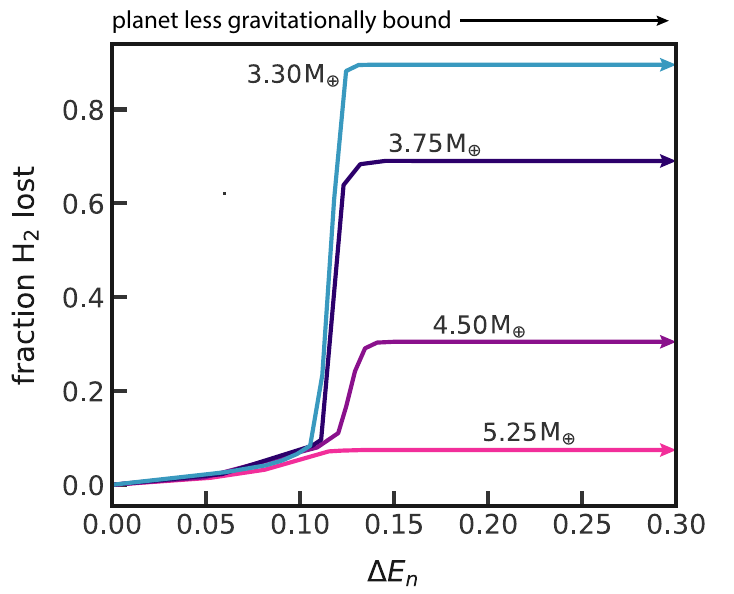}
    \caption{Total hydrogen fraction lost over time versus the normalized energy $\Delta E_{n}$, defined as in Eq. \ref{eq:enorm}, for four miscible planet masses with initial hydrogen mass fractions of 5\%. All planet masses reach their maximum hydrogen fraction lost at a specific $\Delta E_{n}$, and further increases in $\Delta E_{n}$ do not lead to additional mass loss (provided the impact energy does not gravitationally unbind the whole target, i.e. $\Delta E_{n} > 1$). While lower planet masses are almost entirely stripped of their envelopes (light blue), higher mass planets (pink) undergo no or only very little mass loss.}
    \label{fig:hlostve}
\end{figure}

Our results from the previous section show that, once silicate-hydrogen miscibility is accounted for, the radius valley may be preserved even if giant impacts among sub-Neptunes were frequent after the dispersal of the gas disk. Furthermore, our results are also consistent with sub-Neptunes being the progenitors of super-Earths, as previously suggested \citep[e.g.,][]{OwenWu2013, Lopez2013, Ginzburg2018, Gupta2019} rather than super-Earths and sub-Neptunes originating from two compositionally distinct populations \citep[e.g.,][]{Izidoro2022}. 

The reason for this stems from the fact that once silicate-hydrogen miscibility is accounted for, atmospheric mass loss from giant impacts is minimal, typically about 10-20\%, for planets with standard sub-Neptune masses, i.e. $\sim4M_{\oplus}$ (see Fig. \ref{fig:hlostve}). Hence, the sub-Neptune population should keep most of their hydrogen-dominated envelopes during a giant impact phase. Lower-mass planets that ultimately make up the super-Earth population, on the other hand, are frequently stripped of their envelopes during giant impacts after disk dispersal.
Previous work of chemically-inert planets \citep{Biersteker2019,Biersteker2021} suggested that even sub-Neptune planets would be entirely stripped of their atmospheres following giant impacts, thus polluting the radius valley and making it hard to reconcile the existence of a sub-Neptune population with giant impacts after disk dispersal, unless sub-Neptunes are very water-rich and formed as a distinct population from the super-Earths \citep[e.g.,][]{Izidoro2022}.

Recent observations find similar occurrence rates of atmospheric escape in both young (\textless 1 Gyr) and older (\textgreater 1 Gyr) sub-Neptunes \citep[e.g.,][]{Orell2024}. This suggests that atmospheric escape for small, close-in exoplanets is not only due to hydrodynamic escape driven by photoevaporation \citep[][]{OwenWu2013, Owen2017} and core-powered mass loss \citep{Ginzburg2018, Gupta2019, Gupta2020}, which should bias escape detections towards young systems, but that as giant impacts are relatively age-independent \citep[e.g.,][]{Chambers1996,Johansen2012, Pu2015, Rice2023}, they may also play a significant role in driving atmospheric loss.

Furthermore, we emphasize that the mass loss results for the miscible planets presented here are independent of the exact impact energy, provided that the resulting planet mass is high enough to sequester the majority of the hydrogen in the interior. As such, these results can be applied to a wide population of impactor and target planet masses: for example, as shown in Fig. \ref{fig:hlostve}, a collision between a 3.25 $M_{\oplus}$ and 2 $M_{\oplus}$ planet would not undergo any mass loss. Provided that the energy of impact does not entirely gravitationally unbind the planet, the post-impact planet mass can be achieved from any combination of impactor and target masses. We note that while we assume collisions are between two planets with the same hydrogen mass fraction, any initial hydrogen mass fractions can be assumed, provided the post-impact mass and hydrogen mass fraction of the system matches the post-impact conditions for the cases investigated here. A full parameter study that includes investigating a range of equilibrium temperatures, planet masses, and hydrogen mass fractions to determine how this model statistically reproduces the observed exoplanet population is left to future work. 

\subsection{Comparison to Impact Shock-Driven Mass Loss}
So far, we have ignored any mass loss resulting from the impact shock itself. Impact shock-driven mass loss has been investigated for the Earth and terrestrial planets and found to have important consequences both for sculpting the planet's atmosphere and its composition \citep[][]{Genda2003, Genda2005, Schlichting2015}.

Comparing our miscible mass loss results to previous works on mass loss driven by the impact shock \citep[e.g.][]{Schlichting2015}, we find that atmospheric escape from the impact shock itself only becomes dominant for the highest mass planets investigated here (see Fig. \ref{fig:multimasscomp}). Specifically, only as the planet mass increases past $\sim$ 5 $M_{\oplus}$ does atmospheric loss from the impact shock exceed that due to thermal impact heating for our miscible models. However, the total atmospheric loss remains small, with only 5\% loss for 10\% impactor mass fractions. We note that as the impactor mass fraction increases, loss due to impact shock will become more important. Regardless, as the impactor mass increases towards the mass of the target body, we reach the limit in which the impact energy will likely gravitationally unbind the target planet. Finally, it would be interesting to calculate the mass loss due to an impact shock propagating through a fully miscible H$_{2}$-MgSiO$_{3}$ interior in future work. 

\subsection{Discussion of Assumptions and Considerations for Future Work}
\label{sec:limits}
 
\subsubsection{Ideal Gas}
\label{sec:idealgas}
We use an ideal gas to model hydrogen in the envelope. At the binodal, pressures reach several GPa and the temperature is $\sim$ 4000 K. At these conditions, we push the limit for where an ideal gas can accurately describe the behavior of gas within the envelope. We neglect dissociation of H$_{2}$ and ionization. While inclusion of a non-ideal EOS would account for these effects and impact the temperature and density structure of the envelope, we tested our models with non-ideal densities and found similar mass loss results. 

\subsubsection{Inclusion of Iron and Helium}
\label{sec:iron}
We do not include iron in our interior model or helium in our envelope. In order to properly do so, we await a full miscibility phase space for the MgSiO$_{3}$-H$_{2}$-He-Fe system. Inclusion of iron and helium would both act to increase the density of the planet, which would decrease mass loss. Critically, iron does not prevent hydrogen miscibility within sub-Neptunes. Specifically, at temperatures and pressures relevant to the interiors of sub-Neptunes, molecular dynamics simulations find silicate and iron to be fully miscible \citep{Insixiengmay2025}. Additionally, \citet{Young2025} find full miscibility of MgSiO$_3$-H$_2$-Fe within sub-Neptune interiors.  Further, hydrogen has been found to be highly siderophilic at conditions relevant to sub-Neptune interiors \citep{Yuan2020, Tagawa2021}, implying that hydrogen sequestration could be increased for models which properly include iron. This would again reduce mass loss for miscible planets. 

\section{Conclusions}

\label{sec:conclusions}
\begin{figure*}
    \centering
    \includegraphics[trim={0 3.7cm 0 0},clip, width=\linewidth]{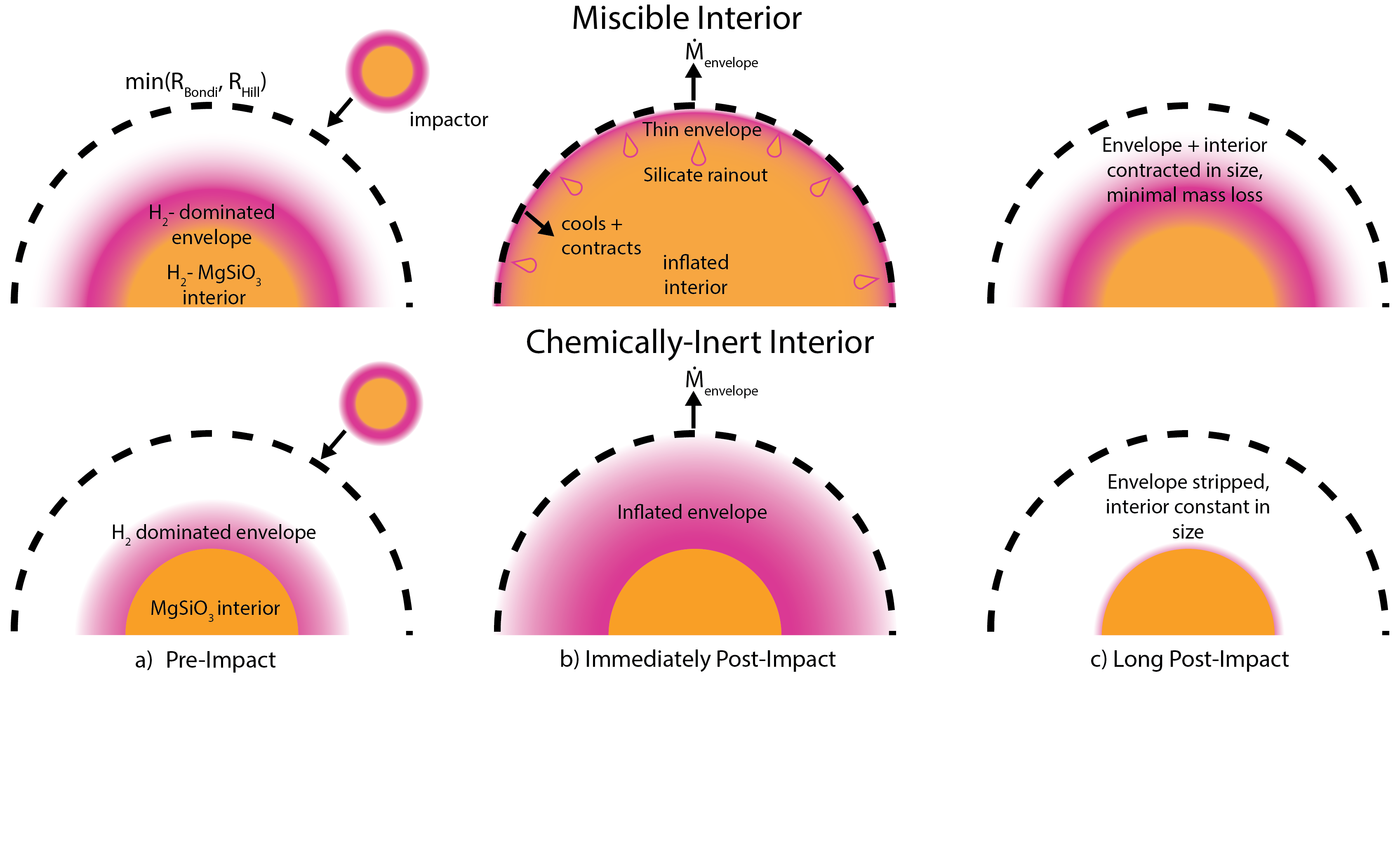}
    \caption{Evolution of a miscible (top panel) and chemically-inert (bottom panel) planet following an impact. In the first column, the miscible and chemically-inert planets initially are cool and have similar interior structures, with a silicate-dominated interior (orange) overlain by a hydrogen-dominated exterior (pink). After an impact, the miscible interior inflates as it sequesters hydrogen, while in the chemically-inert case, the interior size remains constant as the envelope inflates. Mass loss proceeds at the minimum of the Bondi and Hill radii, and is more effective in the chemically-inert case due to the increased hydrogen available in the envelope. This leads to the chemically-inert planet being stripped of its envelope. The miscible planet undergoes less mass loss and is able to exsolve its hydrogen as it cools, maintaining lower mass loss rates. The final state is similar to its initial, pre-impact state.  }
    \label{fig:impcompare}
\end{figure*}

In this work, we have investigated atmospheric loss from sub-Neptunes while accounting for the miscibility between their hydrogen-dominated envelopes and their silicate-rich interiors. Accounting for miscibility produces drastically different giant impact driven atmospheric escape results for sub-Neptunes compared to the previously investigated chemically-inert cases, highlighting the importance of considering the relevant phase equilibria and self-consistently treating the resulting interior structure. 

Specifically, we show that miscible planets undergo less mass loss after a giant impact than chemically-inert planets. This is mainly due to the ability of miscible planets to sequester hydrogen within their interiors. As such, miscible planets post-impact have less available hydrogen mass in their atmospheres, which limits mass loss (see  Fig. \ref{fig:impcompare}). In addition, we find that hydrogen sequestration and atmospheric mass loss for miscible planets is more strongly dependent on planet mass than impact energy, provided that the impact energy does not fully gravitationally unbind the planet. This is due to more massive planets cooling more slowly, allowing for hydrogen to be sequestered within the interior over longer time periods.

Our results also suggest that the radius valley may be preserved, even if giant impacts among sub-Neptunes were frequent after disk dispersal, once silicate-hydrogen miscibility is accounted for. The reason for this stems from the fact that atmospheric mass loss from giant impacts is minimal, typically about 10-20\%, for planets with standard sub-Neptune masses, i.e. $\sim 4 M_{\earth}$. Hence, the sub-Neptune population should keep most of their hydrogen-dominated envelopes during a giant impact phase. In contrast, we find that lower-mass planets that ultimately make up the super-Earth population are frequently stripped of their envelopes during giant impacts after disk dispersal. A full parameter study which investigates a range of equilibrium temperatures, planet masses, and hydrogen mass fractions to determine the extent to which this miscible model is consistent with the observed exoplanet population should be carried out in future work.

\section*{Acknowledgements}

We thank Ed Young for helpful discussions. E.Z. gratefully acknowledges support from the National Science Foundation Graduate Research Fellowship Program under Grant Nos DGE-2034835 and DGE-2444110. Any opinions, findings, and conclusions or recommendations expressed in this material are those of the authors and do not necessarily reflect the views of the National Science Foundation.
JGR gratefully acknowledges support from the Kavli Foundation. This work used computational and storage services associated with the Hoffman2 Cluster which is operated by the UCLA Office of Advanced Research Computing’s Research Technology Group.

\section*{Data Availability}
Data are available upon request.



\bibliographystyle{mnras}
\bibliography{sample701}



\appendix


\bsp	
\label{lastpage}
\end{document}